\documentclass{article}

\usepackage[nonatbib,final]{neurips_2019}

\usepackage{cite}

\usepackage[utf8]{inputenc} %
\usepackage[T1]{fontenc}    %
\usepackage{hyperref}       %
\usepackage{url}            %
\usepackage{booktabs}       %
\usepackage{amsfonts}       %
\usepackage{nicefrac}       %
\usepackage{microtype}      %
\usepackage{graphicx}
\usepackage{subcaption}

\usepackage{wrapfig,lipsum} %
\usepackage{multirow}

\usepackage{xspace}
\newcommand*{\eg}{e.g.,\@\xspace}
\newcommand*{\ie}{i.e.,\@\xspace}
\newcommand*{\cf}{cf.\@\xspace}

\usepackage{xcolor}

\usepackage{colortbl}

\definecolor{c1}{HTML}{369e28} 
\definecolor{c2}{HTML}{5bb961} 
\definecolor{c3}{HTML}{80d494} 
\definecolor{c4}{HTML}{aeeec3} 
\definecolor{c5}{HTML}{ffffe0} 
\definecolor{c6}{HTML}{ffc6b6} 
\definecolor{c7}{HTML}{fd898c} 
\definecolor{c8}{HTML}{e44f6a} 
\definecolor{c9}{HTML}{b02b51} 

\newcommand{\scA}[1]{{\cellcolor{c1} #1}}
\newcommand{\scB}[1]{{\cellcolor{c2} #1}}
\newcommand{\scC}[1]{{\cellcolor{c3} #1}}
\newcommand{\scD}[1]{{\cellcolor{c4} #1}}
\newcommand{\scE}[1]{{\cellcolor{c5} #1}}
\newcommand{\scF}[1]{{\cellcolor{c6} #1}}
\newcommand{\scG}[1]{{\cellcolor{c7} #1}}

\newcommand{\scI}[1]{{\cellcolor{c9} #1}}

\def\ParHead{\vspace*{2mm}\noindent\bf}

\title{SOSD: A Benchmark for Learned Indexes}

\author{%
  Andreas Kipf\thanks{equal contribution} \\
  TUM \\
  \texttt{kipf@in.tum.de} \\
  \And
  Ryan Marcus\footnotemark[1] \\
  MIT CSAIL\\
  \texttt{ryanmarcus@mit.edu} \\
  \And
  Alexander van Renen\footnotemark[1] \\
  TUM \\
  \texttt{renen@in.tum.de} \\
  \And
  Mihail Stoian \\
  TUM \\
  \texttt{stoian@in.tum.de} \\
  \And
  Alfons Kemper \\
  TUM \\
  \texttt{kemper@in.tum.de} \\
  \And
  Tim Kraska \\
  MIT CSAIL\\
  \texttt{kraska@mit.edu} \\
  \And
  Thomas Neumann \\
  TUM \\
  \texttt{neumann@in.tum.de} \\
}

\let\oldmaketitle\maketitle
\renewcommand{\maketitle}{\oldmaketitle\setcounter{footnote}{0}}

\begin{document}

\maketitle

\begin{abstract}
A groundswell of recent work has focused on improving data management systems with learned components.
Specifically, work on learned index structures has proposed replacing traditional index structures, such as B-trees, with learned models.
Given the decades of research committed to improving index structures, there is significant skepticism about whether learned indexes actually outperform state-of-the-art implementations of traditional structures on real-world data.

To answer this question, we propose a new benchmarking framework that comes with a variety of real-world datasets and baseline implementations to compare against.
We also show preliminary results for selected index structures, and find that learned models indeed often outperform state-of-the-art implementations, and are therefore a promising direction for future research.
\end{abstract}

\section{Introduction}
\label{sec:introduction}
There has been a recent surge in proposals to replace traditional database system components, such as query optimizers and index structures, with learned counterparts~\cite{DBLP:conf/cidr/KraskaABCKLMMN19,nashdb,deep_card_est2,neo}.
In particular, learned indexes~\cite{DBLP:conf/sigmod/KraskaBCDP18} raised a lot of attention in the database community.
Learned indexes replace structures such as B-trees with learned models that can accelerate lookups by predicting the position of a sought key in a sorted array (\ie approximating the underlying cumulative distribution function, or CDF).

However,~\cite{DBLP:conf/sigmod/KraskaBCDP18} lacked an open-source implementation and thus left many in the community skeptical that learned models could outperform optimized in-memory data structures~\cite{DBLP:conf/icde/LeisK013,DBLP:conf/sigmod/KimCSSNKLBD10}. Thus, we introduce the \textbf{S}earch \textbf{O}n \textbf{S}orted \textbf{D}ata Benchmark (SOSD), a framework that allows researchers to compare their new (learned) index structures on both synthetic and real-world datasets.
The benchmark is provided as open-source code~\cite{sosd-code} and comes with diverse datasets and highly-optimized baseline implementations.
To the best of our knowledge, our implementation is the first performant and publicly available implementation of the Recursive Model Index (RMI) proposed in~\cite{DBLP:conf/sigmod/KraskaBCDP18}.\footnote{Note, this is a novel implementation, not the implementation used in \cite{DBLP:conf/sigmod/KraskaBCDP18}.}

This extended abstract additionally presents preliminary results from an experimental study using SOSD. We show end-to-end latency measurements and explain the results using performance counters (\eg cache misses, branch mispredictions) for several index structures, including a novel spline-based approach which is trained bottom-up (RMIs are trained top-down). We find that learned index structures can indeed outperform traditional index structures in many scenarios. Finally, we explore the implications of our preliminary findings for practitioners. %

\section{Techniques}
\label{sec:competitors}
\vspace{-2mm}

At a high level, the techniques studied in this work (\cf Table~\ref{tab:competitors}) can be categorized into on-the-fly algorithms, auxiliary indexes, approaches that approximate the cumulative distribution function (CDF) of the data (\ie learned structures), and traditional index structures.
In the context of this work, these techniques are used as in-memory secondary indexes that associate each \emph{key} with a 64-bit \emph{tuple identifier} (TID). %
When queried, each index structure must find all keys (unsigned integers) that qualify under the query predicate, and their associated TIDs.

\begin{wraptable}{r}{7.65cm}
\vspace{-0.2cm}
\caption{Categorization of techniques}
\label{tab:competitors}
\vspace{0.2cm}
\begin{tabular}{@{}lll@{}}
\toprule
category                     & technique           & layout       \\ \midrule
\multirow{3}{*}{on-the-fly}  & BinarySearch (BS)   &  array \\
                             & InterpolationSearch (IS)  &  array \\
                             & TIP                 &  array \\ \midrule
aux. index                   & RadixBinarySearch (RBS)  &  array \\ \midrule
\multirow{2}{*}{approx. CDF} & RadixSpline (RS)    &  array \\
                             & RMI                 &  array \\ \midrule
\multirow{3}{*}{index}       & ART                 & custom       \\
                             & B-tree              & custom       \\
                             & FAST                & custom       \\ \bottomrule
\end{tabular}
\vspace{-0.2cm}
\end{wraptable}

As representatives of \textbf{on-the-fly} algorithms, we studied binary search (BS), interpolation search (IS), and the recently proposed three point interpolation search (TIP)~\cite{DBLP:conf/sigmod/SandtCP19}.
These algorithms operate directly on a sorted array, do not build an index, and do not need to examine the data ahead of query time.
Binary search takes $O(\log n)$ steps to find the sought key, and can cause as many cache misses.
In contrast, interpolation search ``guesses'' the position of the key by interpolating between the current min and max keys.
For dense integer data such as sequential primary keys, this strategy works extremely well and might, in the best case, only incur a single cache miss~\cite{DBLP:conf/damon/Graefe06}.
For highly skewed data, on the other hand, it can degrade to a linear scan.
TIP, an improved interpolation search algorithm, addresses this problem by using linear fractions for interpolation~\cite{DBLP:conf/sigmod/SandtCP19}. %

Similar to on-the-fly algorithms, \textbf{auxiliary indexes} operate on the sorted array, but they additionally build small and thus cache-efficient auxiliary structures.
RadixBinarySearch (RBS) pre-scans the data to populate a flat \emph{radix structure}: an array mapping fixed-length key prefixes to the first occurrence of that prefix in the sorted array. To process a lookup, RBS determines bounds on the key's position using the radix structure. RBS then preforms a binary search in this much smaller range.

\textbf{CDF approximation} algorithms directly estimate a lookup key's position using pre-trained / fitted models, and can thus be considered learned index structures. The RMI builds (a typically 2-level) tree of models (\eg linear, linear spline, log linear) to approximate the CDF, and is constructed top-down as described in~\cite{DBLP:conf/sigmod/KraskaBCDP18}.
In contrast, and similar to~\cite{fiting_tree}, RadixSpline (RS) is built bottom-up, by fitting a linear spline to the CDF~\cite{spline}, and indexes the resulting spline segments in a radix structure. At lookup time, RS locates the corresponding spline segment using the radix structure and performs a linear interpolation between the two spline points.

We also include traditional \textbf{index structures} as baselines: ART~\cite{DBLP:conf/icde/LeisK013} represents radix trees, B-tree is a popular B+-tree implementation~\cite{stxbtree} (STX, v0.9), and FAST~\cite{DBLP:conf/sigmod/KimCSSNKLBD10} is a cache-optimized binary search tree. Notably, all of these structures reorganize the data into a specialized structure.

\section{Search on Sorted Data Benchmark}
\label{sec:evaluation}
\vspace{-2mm}

SOSD is a new benchmarking framework that allows researchers to compare in-memory search algorithms on sorted data.
It is provided as C++ open source code~\cite{sosd-code} that incurs little overhead (8 instructions and 1 cache miss per lookup), comes with diverse synthetic and real-world datasets, and provides efficient baseline implementations.

\begin{figure}
\centering
\includegraphics[width=\linewidth]{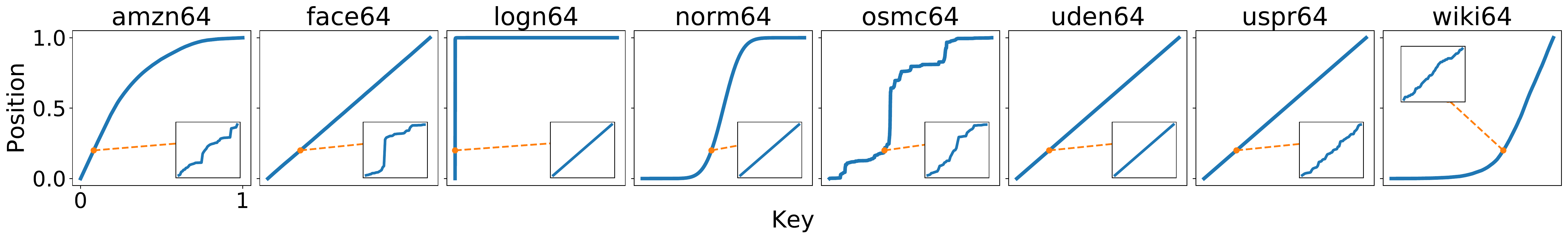}
\caption{CDFs of datasets}
\label{fig:cdf}
\end{figure}

{\ParHead Datasets.}
SOSD currently includes eight different datasets.
Each dataset consists of 200 million 64-bit unsigned integers (keys) with very few duplicates (if at all):
\texttt{amzn} represents book sale popularity data~\cite{amzndata}, \texttt{face} is an upsampled version of a Facebook user ID dataset~\cite{DBLP:conf/sigmod/SandtCP19}, \texttt{logn} and \texttt{norm} are lognormal (0, 2) and normal distributions respectively, \texttt{osmc} is uniformly sampled OpenStreetMap locations~\cite{DBLP:journals/pvldb/PandeyKNK18} represented as Google S2 CellIds~\cite{s2geometry}, \texttt{uden} is dense integers, \texttt{uspr} is uniformly distributed sparse integers, and \texttt{wiki} is Wikipedia article edit timestamps~\cite{wikimedia}.
Figure~\ref{fig:cdf} shows the CDFs of these datasets.
In addition, there are 32-bit versions of all datasets (except \texttt{osmc} and \texttt{wiki}) with similar CDFs.
We use different parameters, (0, 1), for \texttt{logn} in the 32-bit case to reduce the number of duplicates.
When loading the datasets, we generate 64-bit TIDs and store them alongside the keys in row format, although we plan to add support for columnar formats as well.

{\ParHead Queries.}
We perform 10 million equality lookups on a given dataset.
Lookup keys are uniformly chosen from the set of keys.
Note that we disallow hashing implementations (\eg hash tables) since we want to support range-based lookups (\eg lower bound searches).
We ensure that lookup keys have at most 100 matches to limit the impact of result materialization.
Lookups are performed one-at-a-time in a single thread. Our framework verifies each query result for correctness with low overhead (two instructions and a cache miss) and allows for out-of-order execution of lookups.

\begin{table*}
  \small
  \centering
  \caption{Lookup latencies in nanoseconds per lookup. Top approach in each row is green. Yellow indicates 2-3x degraded performance. Red indicates > 3x degraded performance.}
  \label{tab:latency}
  
\begin{tabular}{llllllllll}
\toprule
{} &                ART      &  B-tree  &      BS   &               FAST &                 IS &              RBS  &              RMI    &              RS &      TIP \\
\midrule                                                                                                           
amzn32 &  \color{gray}{n/a} & \scF{529} & \scI{773} & \scA{\textbf{244}} &         \scI{4604} &        \scB{325}  &        \scA{264}    &        \scB{275}    & \scI{731} \\
face32 &  \scA{\textbf{187}}& \scE{524} & \scI{771} &          \scB{229} &         \scI{1285} &        \scB{312}  &        \scB{274}    &        \scD{386}    & \scI{964} \\
logn32 &  \color{gray}{n/a} & \scI{522} & \scI{765} &          \scD{294} &  \color{gray}{n/a} &        \scI{471}  & \scA{\textbf{97.0}} &        \scA{105}    & \scI{744} \\
norm32 &          \scE{191} & \scI{522} & \scI{771} &          \scG{229} &        \scI{10257} &        \scI{355}  & \scA{\textbf{71.7}} & \scA{\textbf{70.9}} & \scI{884} \\
uden32 &          \scD{102} & \scI{521} & \scI{771} &          \scI{228} &\scA{\textbf{39.8}} &        \scI{333}  &       \scA{54.2}    &         \scA{64.2}  & \scG{176} \\
uspr32 &  \color{gray}{n/a} & \scI{524} & \scI{771} &          \scC{230} &          \scG{469} &        \scD{301}  & \scA{\textbf{153}}  &         \scB{200}   & \scF{400} \\
\midrule
size overhead  & 47\% & 16\% & 0\% & 123\% & 0\% & < 1\% & 3\% & < 1\% & 0\% \\
\midrule
amzn64 &  \color{gray}{n/a} & \scC{601} & \scI{804} &  \color{gray}{n/a} &         \scI{4736} &         \scA{387} &           \scA{266} &  \scA{\textbf{288}} & \scI{759} \\
face64 &          \scB{391} & \scI{592} & \scI{784} &  \color{gray}{n/a} &         \scI{1893} &\scA{\textbf{337}} &  \scA{\textbf{334}} &           \scC{461} & \scI{1232} \\
logn64 &          \scE{309} & \scI{597} & \scI{784} &  \color{gray}{n/a} &  \color{gray}{n/a} &         \scI{753} &           \scB{179} &  \scA{\textbf{120}} & \scI{454} \\
norm64 &          \scF{266} & \scD{592} & \scI{785} &  \color{gray}{n/a} &        \scI{10510} &         \scC{405} &  \scA{\textbf{71.5}}&  \scA{\textbf{70.5}}& \scI{862} \\
osmc64 &  \color{gray}{n/a} & \scI{599} & \scG{785} &  \color{gray}{n/a} &        \scI{95076} &         \scI{492} &  \scA{\textbf{402}} &           \scB{437} & \scI{7186} \\
uden64 &          \scB{112} & \scI{592} & \scI{784} &  \color{gray}{n/a} & \scA{\textbf{43.4}}&         \scE{344} &          \scA{54.3} &          \scA{53.9} & \scG{193} \\
uspr64 &          \scC{287} & \scF{591} & \scI{785} &  \color{gray}{n/a} &          \scE{449} &         \scC{313} &  \scA{\textbf{169}} &           \scB{214} & \scF{428} \\
wiki64 &  \color{gray}{n/a} & \scF{608} & \scI{802} &  \color{gray}{n/a} &         \scI{7846} &         \scB{364} &  \scA{\textbf{222}} &  \scA{\textbf{218}} & \scI{1019} \\
\midrule
size overhead  & 25\% & 16\% & 0\% & \color{gray}{n/a} & 0\% & < 1\% & 3\% & < 1\% & 0\% \\
\bottomrule
\end{tabular}

  \vspace{-5mm}
\end{table*}

{\ParHead Results.}
To encourage reproducibility, we use the AWS machine type \texttt{c5.4xlarge} for benchmarking.
Using the Intel Memory Latency Checker~\cite{mlc}, we measured a DRAM latency (LLC miss) of 90\,ns.
\setlength{\columnsep}{0.5cm}
\setlength{\intextsep}{0.2cm}
\begin{wrapfigure}{r}{0.45\linewidth}
\centering
\includegraphics[width=\linewidth]{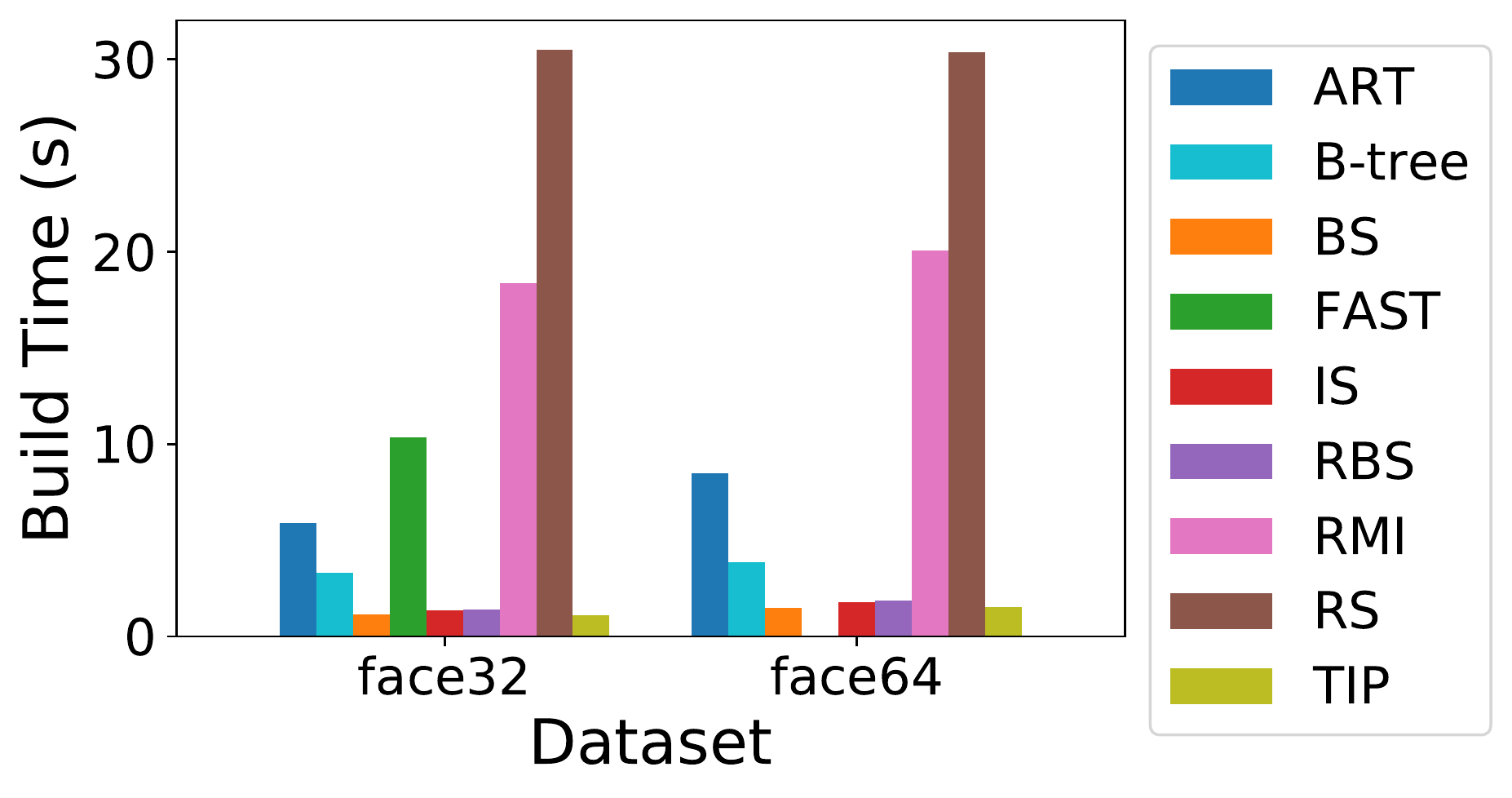}
\caption{Build times}
\vspace{-0.1cm}
\label{fig:build}
\end{wrapfigure}
Table~\ref{tab:latency} shows the lookup performance (in nanoseconds per lookup) for each technique as well as their average size overheads with respect to the data array (the 32-bit array is stored in packed format \ie no padding).
ART performs very well for the 32-bit datasets and even outperforms all others for \texttt{face32}.
Note that ART does not support duplicate keys without modification.
B-tree, BinarySearch, and FAST (which only supports 32-bit keys) are hardly affected by the data distribution.
InterpolationSearch produces very low numbers for the dense data (\texttt{uden32} and \texttt{uden64}) but is heavily affected by skew.
RadixBinarySearch shows consistent improvements over BinarySearch.
The CDF approximators RMI and RadixSpline both have very low lookup latencies, highlighting the benefit learned index structures receive from fitting a model to the data distribution.
TIP improves over the textbook interpolation search in most cases, but is affected by the many steps in \texttt{osmc64} (\cf Figure~\ref{fig:cdf}).

Figure~\ref{fig:build} shows a preliminary analysis of build times.
Note that we have not optimized build phases:
for example, ART could achieve lower build times with bulk loading, and RMI could be improved by building models on samples.
BS, IS, and TIP do not spend any time on build, except for copying the data vector due to our framework design (an approximately two second cost, paid once).
The same holds for RBS, which only needs to perform a scan over the data to build its auxiliary structure.
RS experiences the highest build times for fitting a fine-grained linear spline to the CDF of the data.
However, even without optimizations, the build times of the CDF approximators may be acceptable for many applications.

\begin{figure*}
  \begin{subfigure}{0.32\textwidth}
    \includegraphics[width=\textwidth]{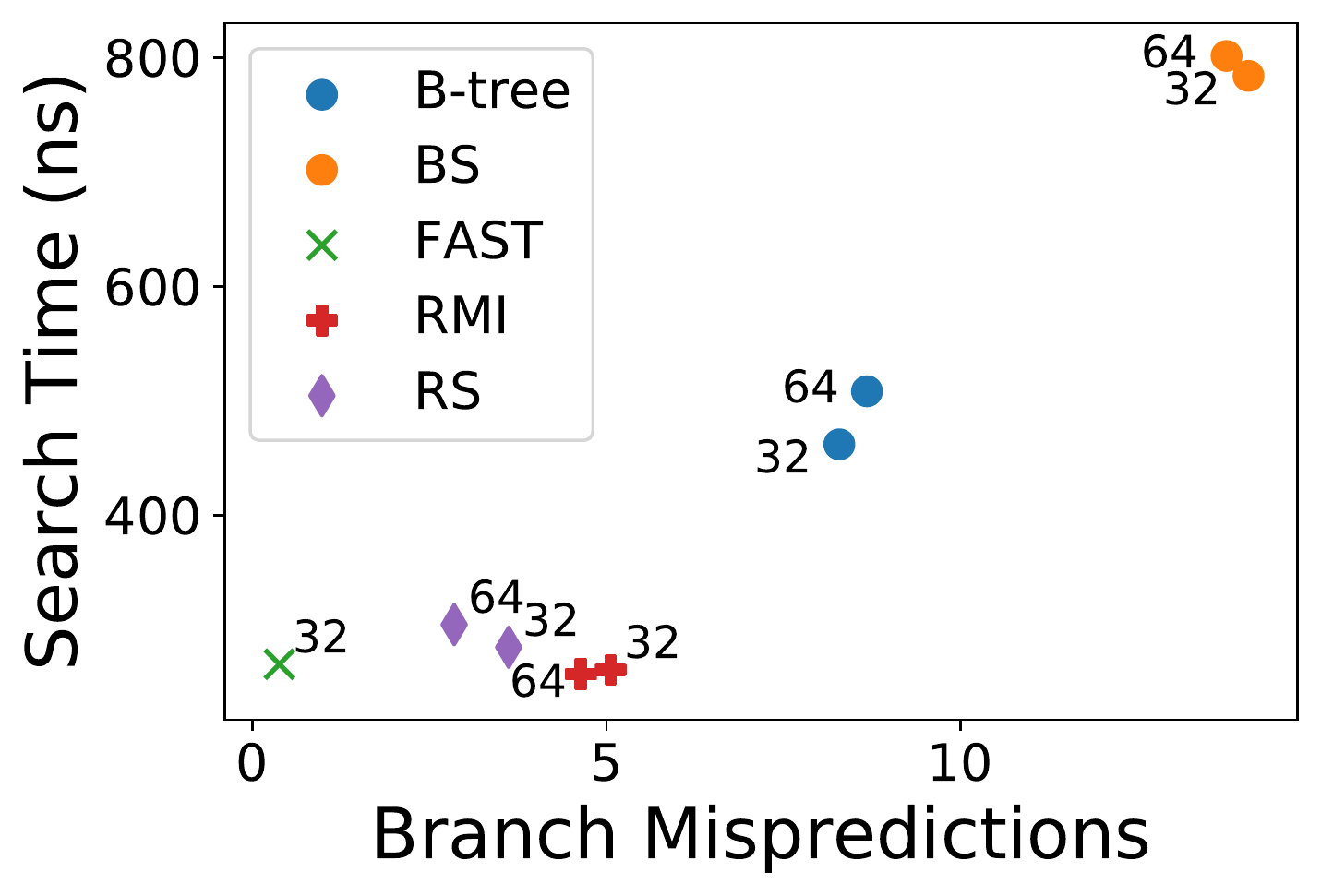}
    \caption{Branch mispredictions}
    \label{fig:branch}
  \end{subfigure}\hfill
  \begin{subfigure}{0.32\textwidth}
    \includegraphics[width=\textwidth]{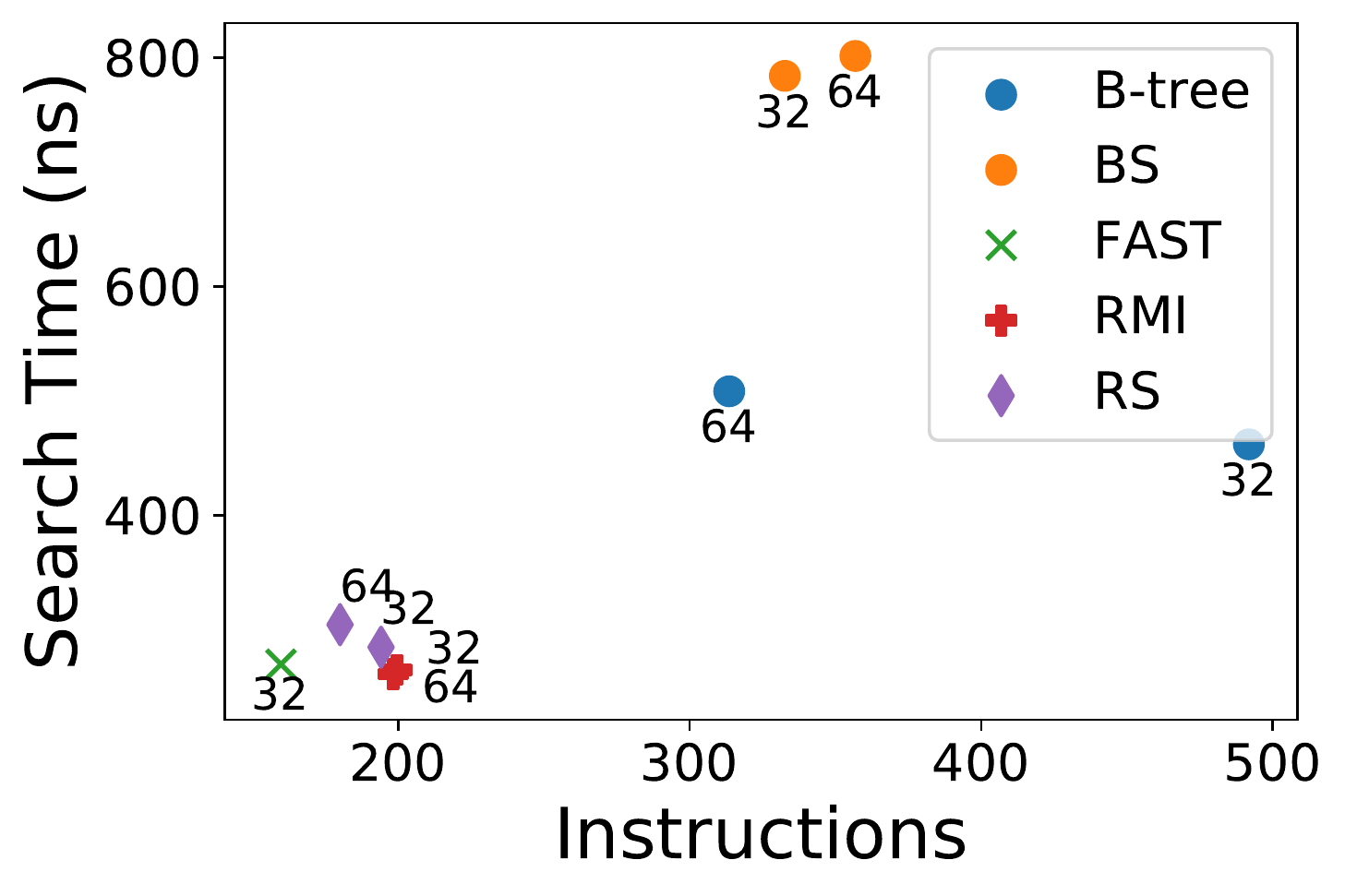}
    \caption{Instructions executed}
    \label{fig:instructions}
  \end{subfigure}\hfill
  \begin{subfigure}{0.32\textwidth}
    \includegraphics[width=\textwidth]{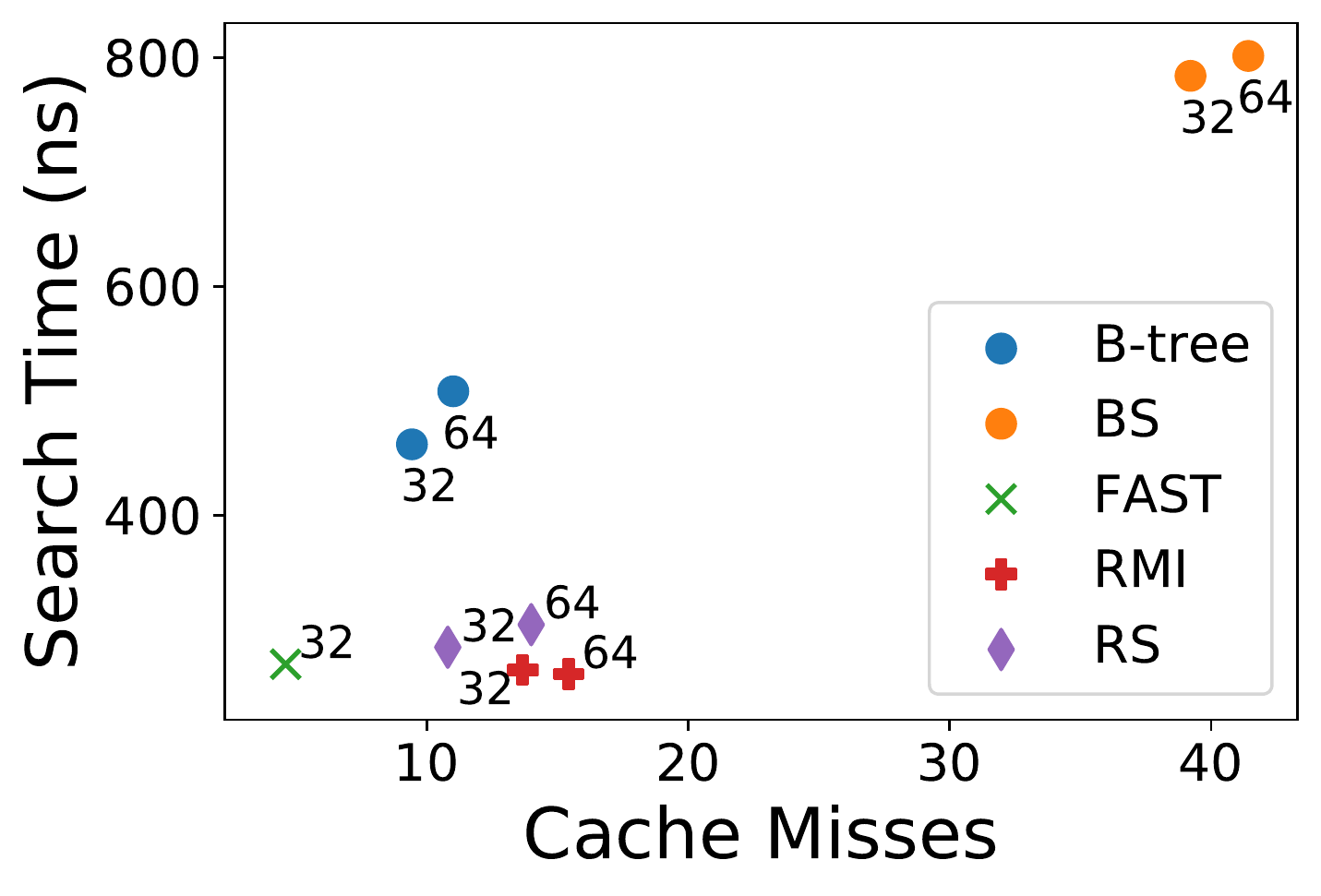}
    \caption{LL cache misses}
    \label{fig:cache}
  \end{subfigure}\hfill
\caption{Performance counter breakdown of amzn32 and amzn64 data}
\label{fig:counters}
\end{figure*}

Figure~\ref{fig:counters} shows performance counters (per lookup) for selected algorithms on \texttt{amzn32} and \texttt{amzn64}, generated automatically by SOSD in \texttt{perf} mode.
Here, we use an Intel Xeon E5-2680 v4 CPU, since AWS does not expose hardware counters.
FAST incurs less than five cache misses per lookup, while RMI and RS incur around ten. However, RMI and RS have similar or superior search times, indicating that analyzing cache misses alone is not sufficient to understand search time. Similarly, neither instructions executed nor branch mispredictions fully explain deviations in search time.
SOSD automatically analyzes branch mispredictions and instructions executed as well.

\section{Takeaways}
\label{sec:conclusions}
\vspace{-2mm}

We have seen that the CDF approximators (RMI, RS) can outperform our baseline implementations.
For the tested datasets, we have shown that simple models are sufficient for efficient learned indexes.
We next explore potential decision points for choosing between different index structures.

For the trivial case of uniform dense integers, non-surprisingly IS is the clear winner.
Otherwise, the optimal search strategy depends on whether a user can afford to manually tune and fit a CDF model, as both RMI and RS require dataset-specific tuning. 
The upside of these models is that they consume very little space compared to our studied index structures. If users can afford this tuning step, we recommend using either RMI or RS.\footnote{RMI and RS both represent learned models -- the primary difference is that RMI is built top-down (starting with the root model), while RS is built bottom-up (starting by fitting linear splines to the data, though other models could be used as well).} 
However, it might be possible to significantly improve the training times of RMI/RS and to do it entirely automatically, without user-intervention. 
Until then, if users cannot afford the training time, we recommend using ART or FAST for 32-bit keys and ART or RBS for 64-bit keys. Unlike ART, RBS can operator directly on a sorted array (i.e., RBS has low space overhead compared to ART).

A current drawback of learned indexes is the lack of support for efficient updates, an arguably important feature for index structures.
However, several recent works have shown progress towards addressing updates~\cite{DBLP:conf/sigmod/WuYTSB19,fiting_tree,DBLP:journals/corr/abs-1905-08898}.
Moreover, it should be pointed out that many of the benchmarked methods (\eg RBS and FAST) also do not support efficient updates.

We plan to extend SOSD with multi-threaded/vectorized lookups, updates, and integration into real database systems to support measuring the impact of these structures on SQL queries. We will additionally compare with recent variants of learned index structures~\cite{pgm-index, DBLP:journals/corr/abs-1905-08898}. We hope SOSD can serve as a platform for testing multi-dimensional index structures, accelerators such as GPUs and FPGAs, and just-in-time compiled custom-tailored data structures for specific queries or datasets.

\clearpage
\bibliographystyle{abbrv}
\bibliography{main,ryan-cites-long}

\end{document}